\documentclass[useAMS,usenatbib]{mn2e}
\usepackage[cp866]{inputenc}
\usepackage[english]{babel}
\usepackage{graphicx}
\usepackage{amsmath}
\usepackage{epsf}

\def\la{\;\raise0.3ex\hbox{$<$\kern-0.75em\raise-1.1ex\hbox{$\sim$}}\;}
\def\ga{\;\raise0.3ex\hbox{$>$\kern-0.75em\raise-1.1ex\hbox{$\sim$}}\;}

\title[Cooling of Akmal-Pandharipande-Ravenhall neutron star models]
{Cooling of Akmal-Pandharipande-Ravenhall neutron star models}

\author[M.~E.~Gusakov, A.~D.~Kaminker, D.~G.~Yakovlev, O.~Y.~Gnedin]
{M.~E.~Gusakov,$^1$\thanks{Send offprint request to: M.~E.~Gusakov}
A.~D.~Kaminker,$^1$
D.~G.~Yakovlev,$^1$
and  O.~Y.~Gnedin$^2$  \\	 
$^1$Ioffe Physical Technical Institute,
Politekhnicheskaya 26, 194021 Saint-Petersburg, Russia,  \\
e-mail: gusakov@astro.ioffe.ru, kam@astro.ioffe.ru, yak@astro.ioffe.ru  \\
$^2$Ohio State University, 760 1/2 Park Street, 
Columbus, OH 43215, USA,  \\
e-mail: ognedin@astronomy.ohio-state.edu}

\date{Released 2005 Xxxxx XX}

\pagerange{\pageref{firstpage}--\pageref{lastpage}} \pubyear{2005}

\def\LaTeX{L\kern-.36em\raise.3ex\hbox{a}\kern-.15em
    T\kern-.1667em\lower.7ex\hbox{E}\kern-.125emX}

\begin{document}

\label{firstpage}

\maketitle

\begin{abstract}
We study the cooling of superfluid neutron stars
whose cores consist of nucleon matter with the
Akmal-Pandharipande-Ravenhall equation of state. 
This equation of state opens the powerful
direct Urca process of neutrino emission 
in the interior of most massive neutron stars.
Extending our previous studies
(Gusakov et al.\ 2004a, Kaminker et al.\ 2005), 
we employ phenomenological
density-dependent critical temperatures
$T_{\rm cp}(\rho)$  
of strong singlet-state proton pairing
(with the maximum
$T_{\rm cp}^{\rm max} \sim 7 \times 10^9$~K
in the outer stellar core) 
and  $T_{\rm cnt}(\rho)$ of 
moderate triplet-state neutron pairing 
(with the maximum $T_{\rm cnt}^{\rm max}  \sim  6 \times 10^8$~K
in the inner core).  
Choosing properly the position of $T_{\rm cnt}^{\rm max}$ we can obtain 
a representative class of
massive  neutron stars 
whose cooling is intermediate between 
the cooling enhanced by the neutrino
emission due to Cooper pairing of neutrons in the absence of 
the direct Urca process
and the very fast cooling provided by the 
direct Urca process non-suppressed by superfluidity.
\end{abstract}

\begin{keywords}
Stars: neutron -- dense matter
\end{keywords}


%
\section{Introduction}

Fundamental properties of supranuclear matter in
the cores of neutron stars, such as the equation
of state and critical temperatures
of baryon superfluidity, are still poorly known.
In particular, they can be studied 
(e.g., \citealt{yp04} and references therein) 
by comparing the theory of cooling isolated
neutron stars with observations of 
thermal radiation  
(e.g.,  \citealt{pzs02}, \citealt{pz03})
from these stars.

In this paper we study the cooling of superfluid neutron stars
with the cores composed of neutrons, protons, electrons,
and muons. We employ model equations of state (EOSs) of neutron star
cores based on the 
EOS of Akmal, Pandharipande \& Ravenhall (1998) (APR, Section 4)
which is currently assumed to be the most elaborated
EOS. It opens the powerful direct Urca process of
neutrino emission \citep{lpph91} 
in neutron stars with masses slightly lower than
the maximum mass. Our cooling models will be based on standard
(nucleon) composition of neutron stars and standard neutrino
processes (no ``exotic'' physics involved). 

The paper extends our previous studies 
(\citealt{gkyg04a} and \citealt{kgyg05}; 
hereafter Paper~I and Paper~II, respectively)
of cooling neutron stars with superfluid nucleon cores,
where the direct Urca process is forbidden. 
In this scenario, observations of cold neutron stars
(e.g., of the Vela pulsar) are explained by the enhanced
neutrino emission due to Cooper pairing of neutrons
in neutron star cores. 
\citet{plps04}, who suggested a similar cooling scenario, 
called it ``Minimal Cooling''.
Note, however, that the realization of
this scheme by Page et al.\ and in Papers I and II
is different (see Paper II for details). 

In Papers~I and II we employed the 
EOS of \citet{dh01}
(the DH EOS, which forbids the direct Urca process in all stars)
and a model of triplet-state neutron pairing
with the maximum of the critical temperature 
$T_{\rm cnt}^{\rm max} \approx 6 \times 10^8$~K 
in the inner stellar core.
This pairing is needed
to produce an enhanced neutrino cooling 
of massive stars and explain the observations  
of neutron stars coldest for their age 
(such as the Vela pulsar). 
In Paper I we assumed also strong singlet-state 
proton pairing in the outer stellar core, 
with $T_{\rm cp}^{\rm max} \ga 5 \times 10^9$~K.
It slows down the cooling 
of low-mass neutron stars and enables us to interpret 
the observations of stars 
warmest for their age (e.g., RX J0822--4300 or PSR B1055--52). 
In Paper~II we included the effects of 
surface (accreted) envelopes of light elements. 
They  allowed us to explain the observations of  
warmest stars assuming weaker proton
pairing ($T_{\rm cp}^{\rm max} \ga 10^9$~K).

Now we will add the effect of the direct Urca process
in most massive stars in accordance with the APR EOS. 
We will use similar (but slightly different) models
for strong proton pairing 
and moderate neutron pairing in 
the stellar core.
These superfluid models 
are sufficient to explain current observations of
all cooling neutron stars without opening the direct Urca process. 
However, most massive stars predicted by the APR EOS 
can be much colder than the stars observed today
and, we will focus on their properties. 
For simplicity, we will neglect the effects of
accreted envelopes.
They are unimportant for massive neutron stars;
we could easily include them as in Paper II. 
 
\section{Observations}
\label{observ}

Table~1 summarizes the observations of isolated (cooling)
middle-aged neutron stars ($10^3$~yr $ \la t \la 10^6$ yr),
whose thermal surface radiation has been detected or constrained.
We present the estimated stellar ages $t$ and the
effective surface temperatures $T_{\rm s}^\infty$
(redshifted for a distant observer).
The data are described in Paper~I and 
slightly  corrected in Paper~II.

\begin{table*}   
\caption[]{Observational limits on surface temperatures of isolated
neutron stars}
\label{tab:observ}
\begin{center}
\begin{tabular}{@{}lcccl}
\hline
\hline
Source & $t$ [kyr] & $T_{\rm s}^\infty$ [MK] &  Confid.\ of $T_{\rm s}^\infty$
& References   \\
\hline
\hline
PSR J0205+6449\   (in 3C 58)  & 0.82    & $<$1.02$^{~b)}$  &  99.8\%   & 
\citet{slane04b}    \\
Crab               &    1    & $<$2.0$^{~b)}$   &  99.8\%    & 
\citet{weiss04}  \\
RX J0822--4300     & 2--5    & 1.6--1.9$^{~a)}$ & 90\% & 
\citet{ztp99}   \\
1E 1207.4--5209        & 3--20 & 1.4--1.9$^{~a)}$ & 90\% & 
\citet{zps04} \\
RX J0007.0+7303  (in CTA 1)  & 10--30  &     
$<$ 0.66$^{~b)}$ & --   &
\citet{hgchr04} \\ 
Vela               & 11--25  & 0.65--0.71$^{~a)}$ & 68\% & 
\citet{pzs02} \\
PSR B1706--44       & $\sim$17 & 0.82$^{+0.01}_{-0.34}$$^{~a)}$ & 68\% &
\citet{mcgwn04}  \\
PSR J0538+2817       & $30 \pm 4$ & $\sim 0.87$$^{~a)}$ & -- &
\citet{zp04} \\
%
%
Geminga            & $\sim$340 & $\sim 0.5$$^{~b)}$ & -- &
\citet{kpzr05}  \\
RX~J1856.4--3754     & $\sim$500 & $<$0.65  & -- & see Paper I \\
PSR~B1055--52       & $\sim$540 & $\sim$0.75$^{~b)}$ & -- &
\citet{pz03}  \\
RX J0720.4--3125   & $\sim 1300$ & $\sim 0.51$$^{~a)}$  & -- &
\citet{mzh03} \\
\hline
\end{tabular}
\end{center}
\medskip
$^{a)}$ Inferred using a hydrogen atmosphere model\\
$^{b)}$ Inferred using the black-body spectrum\\
\end{table*}
 
The surface temperatures of some 
sources (labeled by $^{a)}$)
have been obtained by fitting their thermal radiation
spectra with hydrogen atmosphere models. Such models are
more consistent with other information on these sources
(e.g., \citealt{pz03}) 
than the blackbody model. For other sources 
(e.g., for Geminga and PSR B1055--52,
labeled by $^{b)}$) we present the values
of $T_{\rm s}^\infty$ inferred using the blackbody spectrum
because this spectrum is more consistent for these 
sources. The surface temperature of RX J1856.4--3754 
is still uncertain. Following Paper~I we adopt the upper
limit $T_{\rm s}^\infty < 0.65$~MK. 
Finally, $T_{\rm s}^\infty$
for RX J0720.4--3125 is taken from \citet{mzh03},  
who interpreted the observed spectrum 
with a model of a hydrogen atmosphere
of finite depth.   
             
Note also new results by \citet{kpzr05} for Geminga
presented in Table~1. 
As in Papers~I and II we
retain $20\%$ errorbars  for this source. 

\section{Cooling without the direct Urca process}
\label{physics}

We will simulate neutron star cooling 
using our general relativistic 
cooling code \citep{gyp01} which 
calculates cooling curves
(the dependence of  
$T_{\rm s}^\infty$ on the stellar age $t$).
At the initial  
cooling stage ($t \la 100$ yr) a star cools via neutrino emission but
the stellar interior is non-isothermal. 
At the next  
stage ($10^2$~yr $ \la t \la 10^5$ yr) the neutrino emission
dominates but the interior is isothermal.  
Later ($t \ga 10^5$ yr)  
the star cools mainly via the surface photon emission.

\begin{figure*}   
\epsfysize=80mm
\centering
\hspace{-2cm}
\epsffile[18 145 585 710]{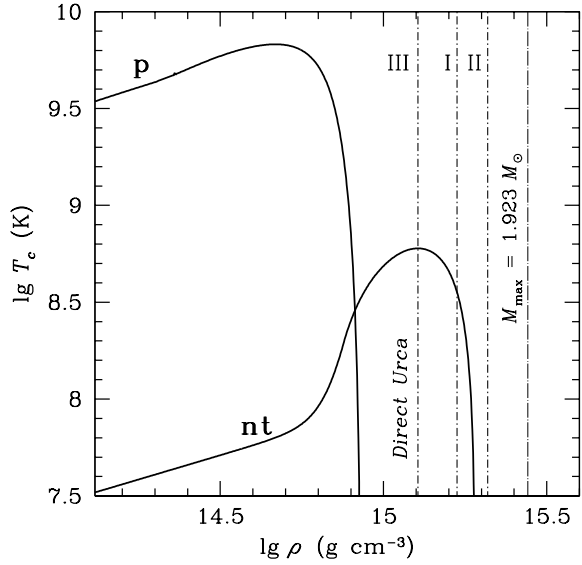}
\vspace{-0.2cm} 
\caption{
Density dependence of the critical temperatures 
for model p 
of singlet-state proton pairing and  
model nt of triplet-state neutron 
pairing in a neutron star core. 
Three vertical dot-and-short-dashed lines 
indicate the thresholds   
of the direct Urca process
for the EOSs  
APR~I, II, and III discussed in Section 4.   
The vertical dot-and-long-dashed line shows 
the central density 
of a maximum-mass neutron star for APR~I;
two other maximum-mass lines are 
almost the same. 
}
\label{fig1}
\end{figure*}

For the adopted EOSs,
all constituents of dense matter 
(nucleons, electrons, and muons) 
exist everywhere in the stellar core, except for muons
which appear at $\rho \sim  2.0 \times 10^{14}$
g~cm$^{-3}$.  
As in Paper~I  
we use the relation 
between the effective surface temperature
and the temperature at the bottom
of the heat-blanketing iron envelope 
calculated  
by \citet{pycg03}.

\begin{figure*}   
\epsfysize=110mm
\hspace{1mm}
\epsffile[18 145 585 710]{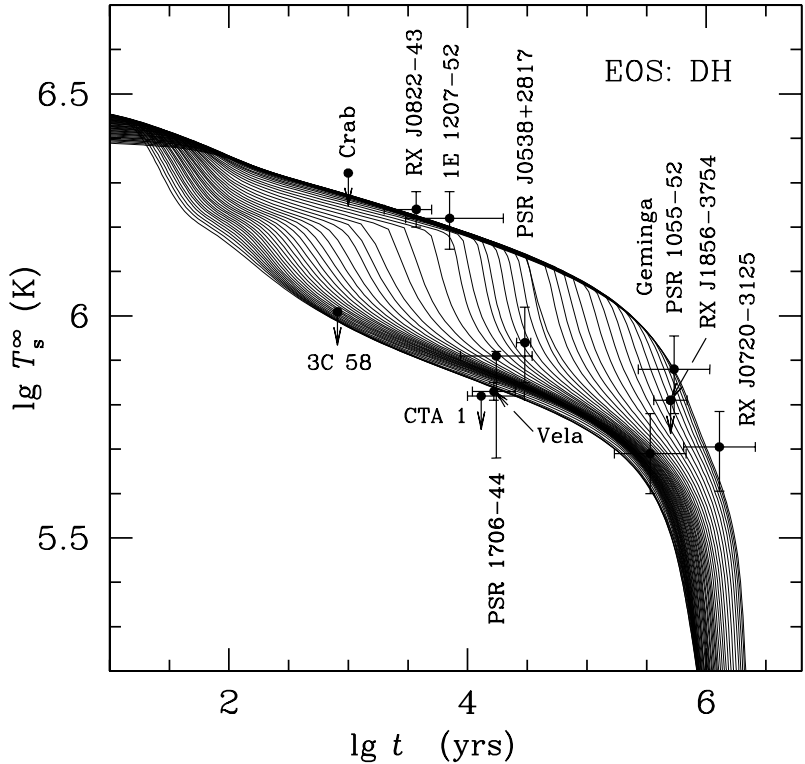}   
\caption{
Observational limits on surface temperatures  
of twelve isolated neutron stars (Table~1)  
compared with theoretical cooling 
curves  calculated for the DH EOS (no direct Urca process)  
and for models p and nt
of nucleon pairing (Fig.\ \protect{\ref{fig1}}). 
Fifty three  
curves refer to neutron stars of masses     
from 1.00 to 2.04~$M_\odot$  
(from top to bottom, with the step $0.02~M_\odot$).  
}
\label{fig2}     
\end{figure*}
  
Neutron star cooling
is strongly affected by superfluidity of nucleons 
in stellar interiors. Any superfluidity
is characterized by a density-dependent profile of critical 
temperature $T_{\rm c}(\rho)$.
In principle, an EOS and a superfluidity model, employed
in cooling simulations, have to
be obtained from the same nuclear Hamiltonian. However,
such calculations are currently absent and one usually
treats the EOS and superfluid properties as independent.
Although this approach is simplified it sounds reasonable at the
present stage of the investigation because the EOS is the bulk
property of the matter (determined by the entire Fermi see
of baryons) while superfluidity is a Fermi-surface phenomenon
and can be, therefore, relatively independent.
Microscopic theories predict 
two main types of superfluidity inside the neutron star  
core: singlet-state proton pairing 
($T_{\rm c}=T_{\rm cp}$) 
and triplet-state neutron pairing  
($T_{\rm c}=T_{\rm cnt}$).   
These theories give
a large scatter of critical temperatures depending on
a nucleon-nucleon interaction model and a many-body theory
employed  (e.g., \citealt{ls01}, \citealt{yls99}; 
see also recent papers by 
\citealt{sf04}, \citealt{tt04}, 
\citealt{zuo04}, \citealt{tmc04}).
As in Papers~I and II, we
will treat $T_{\rm cp}(\rho)$ and $T_{\rm cnt}(\rho)$
as phenomenological functions of $\rho$ 
which do not contradict  
some microscopic calculations. 
These functions
could be (in principle)  
constrained by comparing theoretical cooling
curves with observations.

Superfluidity of nucleons 
affects the nucleon heat capacity  
and suppresses neutrino processes
involving superfluid nucleons: 
Urca and nucleon-nucleon bremsstrahlung processes 
(as reviewed, e.g., by  \citealt{yls99}). 
Moreover, superfluidity induces 
an additional neutrino emission
mechanism associated with Cooper pairing of 
nucleons (Flowers, Ruderman \& Sutherland 1976). 
All these effects
of superfluidity are included into our cooling code
in the same manner as in Papers I and II.

In Fig.\ \ref{fig1}
we plot two models of nucleon 
pairing 
employed
in the present paper,  
model p for strong singlet-state 
proton pairing (with the peak
of $T_{\rm cp}(\rho)$ approximately 
equal
to $T_{\rm cp}^{\rm max} \approx  6.8 \times 10^9$~K); 
and model nt for moderate
triplet-state neutron pairing   
($T_{\rm cnt}^{\rm max} \sim 6 \times 10^8$~K).
Models p and nt   
slightly differ from similar models p1 and nt1 
in Papers~I and II.
Note that the curves $T_{\rm cp}(\rho)$ and 
$T_{\rm cnt}(\rho)$ are almost 
insensitive to the employed EOS:
DH (Papers~I and II) and APR~I, II, and III 
(Section~4). 
Our pairing model nt
seems specific (shifted to too high densities $\rho$
at which neutron pairing dies out
according to many microscopic calculations).
However similar models have been obtained from some
microscopic theories (e.g., see the curve $m^*=0.73$
in Fig.\ 1 of \citealt{tt97}).

Figure 2 illustrates the cooling scenario of Paper I
(to be compared with new scenarios).
We plot 
a family of cooling curves calculated
for the DH EOS, which
forbids the direct Urca process 
in all stable neutron stars
($M \leq M_{\rm max}=2.05 M_\odot$).
The curves refer to neutron stars with
(gravitational) masses $M$ from 1.00 to 2.04~$M_\odot$
(from top to bottom). 

The curves explain all the data from Table 1.
The coldest objects 
can be interpreted as 
most massive neutron stars 
with moderate neutron
pairing in the inner cores (at densities
$\rho \ga 8 \times 10^{14}$ g cm$^{-3}$).
When the internal temperature of a massive
star falls below
$T_{\rm cnt}^{\rm max}$, the cooling is noticeably enhanced by
the neutrino emission due to Cooper pairing of neutrons.
This enhancement is sufficient to explain the observations
of neutron stars coldest for their age, 
particularly, PSR J0205+6449, 
RX J0007.0+7303, 
the Vela and Geminga pulsars. 
Note that the data on
RX J0007.0+7303 are explained in this way only marginally. However,
taking into account that real uncertainties of $T^\infty_{\rm s}$ and $t$
can be larger than the adopted ones we can accept this explanation
as successful.
All stars with the central density 
higher than the density 
at which neutron pairing dies out 
($\rho_{\rm c} \ga 2 \times 10^{15}$~g~cm$^{-3}$
in Fig.\ \ref{fig1})
cool nearly as fast as the star with 
$M=M_{\rm max}$. This  
gives the lower dense bunch of cooling curves 
of massive neutron stars (with $\rho_c \ga 2 \times 10^{15}$ g cm$^-3$) 
in Fig.\ \ref{fig2}.    

There is the other (upper) dense bunch of
cooling curves  
of low-mass stars ($M \la 1.1\, M_\odot$). 
The central densities of such stars
are $\rho_{\rm c} \la 8 \times 10^{14}$~g~cm$^{-3}$.
Proton pairing is strong 
($T_{\rm cp}(\rho) \ga  3 \times 10^9$~K) 
everywhere in their cores
(Fig\ \ref{fig1}).
It completely suppresses all neutrino processes 
involving protons and substantially slows down the cooling.

Finally, there
is a class of medium-mass neutron stars 
which show intermediate cooling. Their cooling curves
fill in the space between the upper dense bunch 
of cooling curves (low-mass stars)
and the lower dense bunch  (high-mass stars). 
We can treat PSR B1706--44, PSR J0538+2817,
and RX J1856.4--3754 as medium-mass stars.   

Although this scenario is consistent with all
current observations, 
colder stars could be detected in the future.
An indirect evidence for 
the existence of such stars 
is given by a non-detection of
compact central objects in four supernova remnants
\citep{kaplan04}. If neutron stars are present there,
they should be really cold. It would be impossible
to explain them with the cooling scenario presented
in Fig.\ 2, but they could be interpreted adopting
the APR EOS.

\section{Cooling of APR neutron stars}
\label{Extmincool}

\subsection{Equations of state}
\label{EOS}

Let us employ three 
EOSs of dense matter, APR~I, APR~II, and APR~III. 
They represent a useful parameterization of the
well-known 
EOS obtained by \citet{apr98} 
(model Argon V18 + $\delta v$ + UIX$^*$).
This parameterization was proposed by \citet{hhj99}. 
We use the same parameters for 
the compressional modulus of the nucleon matter 
and vary only the parameter $\gamma$ 
in the expression for symmetry energy, 
$S(n_{\rm b})=S_0 \, (n_{\rm b}/n_{\rm 0})^\gamma$, 
where 
$S_0=32$~MeV;
$n_{\rm b}$ is the baryon number density, and
$n_0=0.16$ fm$^{-3}$ is the baryon number density
in saturated nuclear matter.  
Specifically, we have taken $\gamma$=0.6,
0.575, and 0.643 for the APR EOSs I, II, and III, respectively.
These EOSs differ mainly in 
the threshold densities
for opening 
the direct Urca process  
(see Fig.\ 1).

\begin{figure*}
\vspace{0.5cm}
\epsfysize=80mm
\epsffile[18 145 569 418]{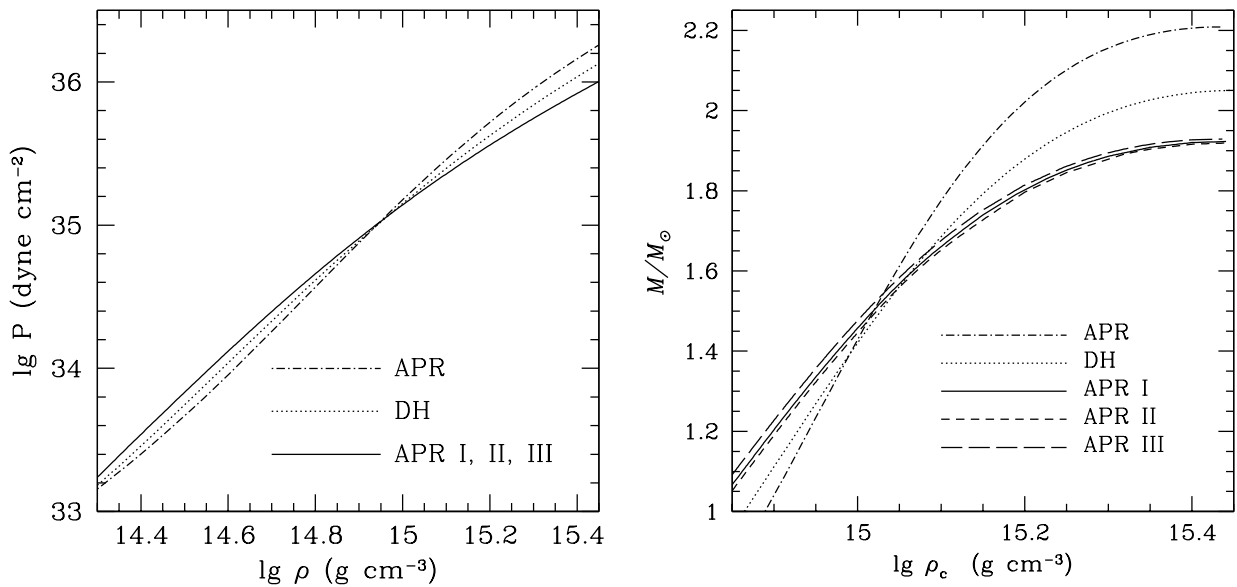}   
\vspace{-0.5cm}
\caption{ 
Pressure -- density (left) and 
mass -- central density (right)
diagrams for five EOSs.   
The APR EOS was suggested 
by \protect{\citet{apr98}}.
The EOSs APR~I,  II, and  III  refer to 
different values of $\gamma$ 
(see Table~2) in the
parameterization proposed by  
\protect{\citet{hhj99}}.  
The DH EOS was derived by  
\protect{\citet{dh01}}. 
}
\label{fig3}    
\end{figure*}

Some properties of the EOSs and associated neutron star
models are illustrated in Figs.\ \ref{fig1} and \ref{fig3}
and in Table 2. For comparison, in Fig.\ \ref{fig3} and Table 2
we include also the data on the DH EOS \citep{dh01} 
and the original APR EOS \citep{apr98}. 
The APR EOS corresponds to the nucleon matter with  
the appearance of neutral 
pion condensation.    
To plot this EOS 
in Fig.\ \ref{fig3} we use 
the parameterization    
proposed by \citet{hp04}, 
with the values of fit parameters
provided by A.Y.\ Potekhin (2004, private communication).

Table 2 gives
masses $M$, central densities $\rho_{\rm c}$, 
and circumferential radii $R$ of two
stellar configurations.
The first configuration is the most massive stable neutron star.
The values of $M_{\rm max}$
indicate that all five EOSs are moderately stiff
and similar.    
The second configuration refers 
to the onset of the direct Urca process,
($\rho_{\rm c}=\rho_{\rm D}$, $M=M_{\rm D}$);
it is absent for
the DH EOS.
The APR~III EOS implies larger symmetry energy
at supranuclear densities 
(larger $\gamma$) and larger 
proton fraction than the EOSs APR I and II. Accordingly,
the direct Urca process opens at lower density for APR III.

\begin{table*}   
\begin{center}
\caption[]{Neutron star models for the EOSs DH, APR I, II, III, and APR}
\label{tab:models}
\begin{tabular}{@{}llccccc}
\hline
Model & Main  & DH & APR & APR~I &   APR~II &   APR~III   \\   
      & parameters  & & & $\gamma= 0.6$ & 0.575 & 0.643  \\ 
\hline
\hline
Maximum  & $M_{\rm max}/M_\odot$ & 2.05 &  2.20  & 1.923   &  1.919   
&  1.929  \\
mass     & $\rho_{\rm max}/10^{15}$~g cm$^{-3}$  
& 2.86  & 2.74 & 2.759  &  2.774  &  2.731   \\
model    & $R$~km  &  9.99  & 9.99  & 10.32  & 10.28  &  10.39    \\
\hline
\hline
Direct Urca &  $M_{\rm D}/M_\odot$ &  -- & 2.01 & 1.829 &  1.891 & 1.685   \\     
threshold   &  $\rho_{\rm D}/10^{15}$~g cm$^{-3}$ 
        & --& 1.56  & 1.680&   2.084 &  1.275 \\ 
model       &  $R$~km  & --& 10.95 & 11.27 & 10.83  & 11.83  \\
\hline
\end{tabular}
\end{center}
\medskip
\end{table*}

Figure \ref{fig3} shows the pressure-density 
(left) and the mass-central density 
(right) diagrams for all five  
EOSs. According to the left panel, all EOSs 
are similar. The $P(\rho)$ curves intersect
in one point, 
$\rho \simeq 8.5 \times 10^{14}$~g~cm$^{-3}$.
However, small differences in the EOSs 
lead to noticeable differences in 
neutron star configurations, particularly, in
the mass-density relations $M(\rho_{\rm c})$ (the right panel). 
Corresponding proton fractions  and
direct Urca thresholds are also different (Table~2 and Fig.\ 1).

\subsection{Beyond the direct Urca threshold}
\label{inter}
 
In contrast to the cooling scenarios in Papers~I and II 
we use the EOS which opens the direct 
Urca process in massive neutron stars.
Moreover, contrary to our older 
cooling scenarios (e.g.,  \citealt{kyg02},   
\citealt{yp04} and references therein)  
the direct Urca process is now open
at densities at which
proton superfluidity dies out (Fig.\ \ref{fig1}), 
in agreement with some recent
calculations of proton pairing in dense
matter (e.g., \citealt{tt97, tt04},  \citealt{zuo04}).   

\begin{figure*}   
\epsfysize=110mm
\hspace{1mm}
\epsffile[18 145 585 710]{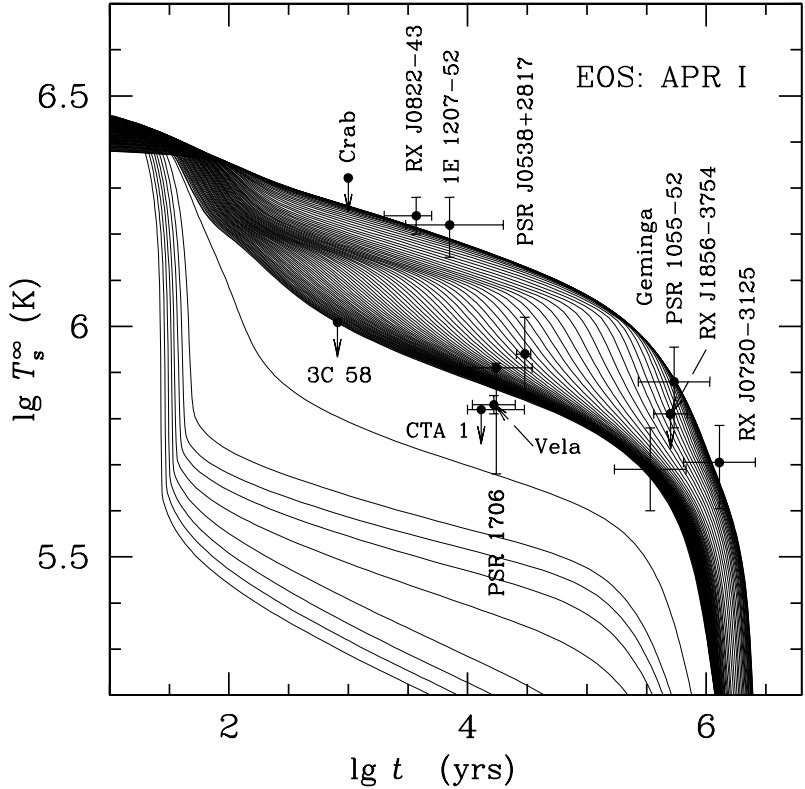}   
\caption{
Same as in Fig.\ \protect{\ref{fig2}} but for
the EOS APR~I.  
Ninety two cooling 
curves refer to neutron stars of masses    
from 1.01 to $M_{\rm max}=1.92~M_\odot$   
(with the step $0.01 M_\odot$). 
The sharp boundary between 
densely and rarely covered domains 
corresponds to the direct Urca
threshold, $M_{\rm D}=1.83 M_\odot$
(Table~2).  
}
\label{fig4}     
\end{figure*}

Figure \ref{fig4}
shows a family of cooling curves
similar to that in Fig.\ \ref{fig2}. It is calculated 
for the same models
of proton and neutron pairing
(p and nt, Fig.\ 1) but adopting the APR I EOS.
The curves show the cooling  
of neutron stars with masses 
from 1.01 to $M_{\rm max}=1.92~M_\odot$.    
The density of cooling curves reflects 
the probability to detect such stars provided the 
mass distribution of stars
is flat in the indicated mass range
(and observational
selection effects are absent).
Stars with $M \leq M_{\rm D}=1.83 \, M_\odot$
would have temperatures nearly in the same range as in Fig.\ \ref{fig2}
(see Section 3).  
By opening the direct Urca we get a new but rare population of
very cold neutron stars ($M_{\rm D} \leq M \leq M_{\rm max}$). 
Taking into account the
difficulties to detect such stars, the chances to observe
them are really low. Notice that if the EOS would allow 
higher $M_{\rm max}$ we could obtain a new dense bunch
of curves for very cold stars with $M \approx M_{\rm max}$.

\begin{figure*}   
\vspace{0.5cm}
\epsfysize=80mm
\epsffile[18 145 569 418]{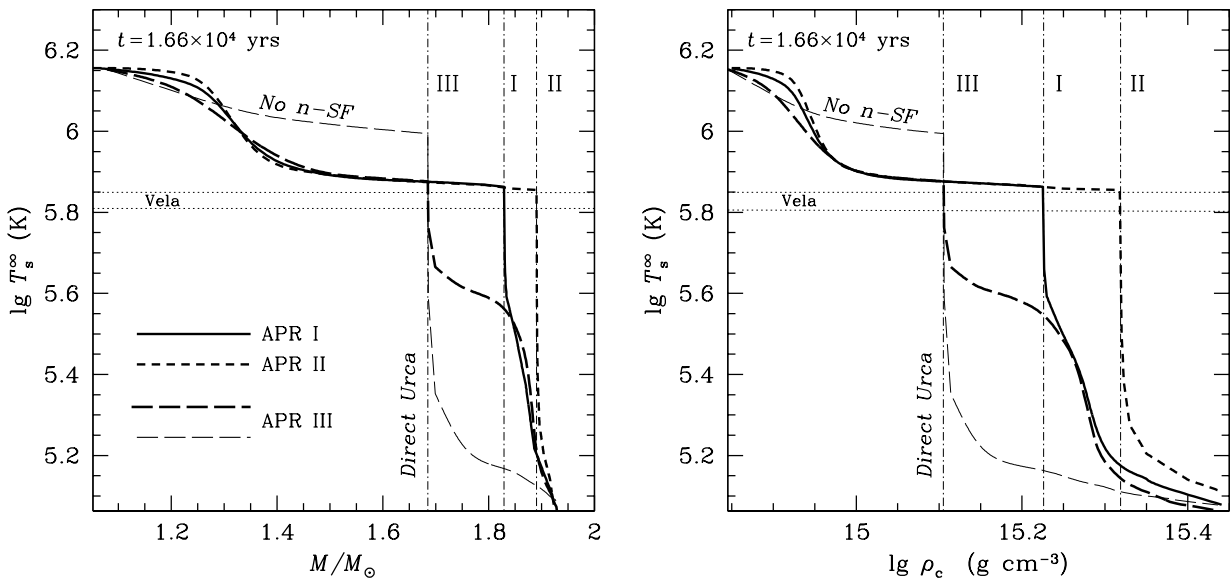}   
\vspace{-0.3cm}
\caption{ 
Surface temperatures of neutron stars   
of the age of Vela pulsar,
\protect{$t=1.66 \times 10^4$}~yr, 
versus stellar mass (left) 
and central density (right)  
for three EOSs (Table~2):  
APR~I (solid lines), APR~II (short dashes), 
and APR~III (thin and thick long dashes), 
assuming either  
combined proton  
and neutron (p and nt) pairing in stellar cores 
(thick lines)
or only proton pairing alone 
(\textit{No} n-\textit{SF}; thin lines).    
The vertical dot-and-dashed lines,  
denoted by I, II, and III,
show the threshold masses $M_{\rm D}$ (left) 
or densities $\rho_{\rm D}$ (right)
for the direct Urca process. 
Horizontal dotted lines 
are observational 
limits on the Vela's surface temperature.  
}
\label{fig5}     
\end{figure*}

The same effects
are demonstrated in Fig.\ \ref{fig5}. It shows  
$T_{\rm s}^\infty$ versus $M$ 
(left) 
or $\rho_{\rm c}$
(right) for stars of
the Vela pulsar age, $t=1.66 \times 10^4$ yrs.
We present the curves for superfluid stars 
with the EOSs APR~I, II, and III. 
The horizontal dotted lines give the observational 
limits of $T_{\rm s}^\infty$  for the Vela pulsar.

One can see a very sharp fall of $T_{\rm s}^\infty$
in the narrow mass range 
from $M=M_{\rm D}$ to $M=M_{\rm D} + 0.01 M_\odot$ 
for the APR~I EOS and especially for APR~II.
This fall is naturally associated with
the strong enhancement of the neutrino luminosity
by the direct Urca process.

The case of ARP~III is different.   
As seen in Fig.\ \ref{fig1}, the 
threshold of opening the direct Urca process
is placed now in a wide region near the maximum
of the critical temperature $T_{\rm cnt}(\rho)$.
When the internal stellar temperature falls
noticeably below 
$T_{\rm cnt}(\rho)$ in some region of  
$\rho > \rho_{\rm D}$, the 
suppression of the direct Urca process 
by neutron superfluidity
becomes important and smears out the dependence
of $T_{\rm s}^\infty$ on $M$. 
The similar effect of partial suppression of the
direct Urca process by superfluidity was discussed by \citet{pa92}. 
Evidently, the level of this suppression is regulated by the
strength and density range of neutron pairing and by
the threshold density of the direct Urca process.
Were neutron superfluidity absent
(the thin long-dashed line),
the dependence would be as sharp as for the EOSs ARP~I and~II.

The dependence of $T_{\rm s}^\infty$  on  $\rho_{\rm c}$
(the right panel of Fig.\ \ref{fig5}) 
is similar to the dependence on $M$.
Note only much smoother dependence of
$T_{\rm s}^\infty$ on $\rho_{\rm c}$ 
in the lower parts of the curves;  
small variations of $M \la M_{\rm max}$
correspond to larger variations of $\rho_{\rm c}$ (Fig.\ \ref{fig3}).

\begin{figure*}   
\vspace{0.5cm}
\hspace{1mm}
\epsfysize=110mm
\epsffile[18 143 580 700]{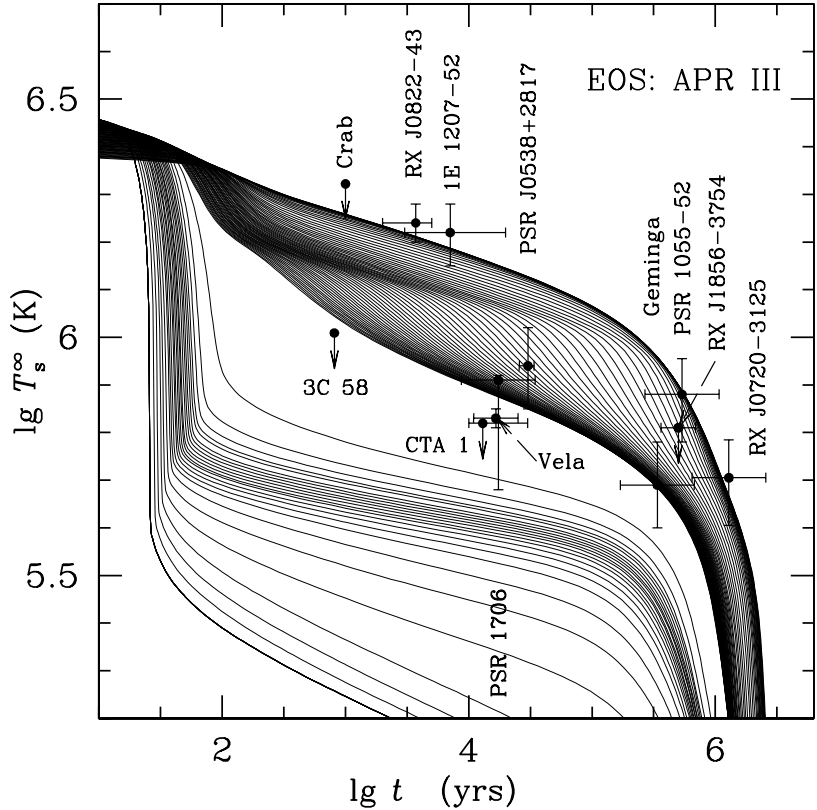}  
\vspace{-0.1cm}
\caption{ 
Same as in Figs.\ \protect{\ref{fig2}} and \protect{\ref{fig3}} 
but for the   
APR~III EOS. Ninety three thin solid  
curves refer to neutron stars of masses    
from 1.01 to $M_{\rm max}=1.93\,M_\odot$ (with the  
step $0.01\, M_\odot$).
The lower boundary of the
densely covered domain 
refers to the direct Urca threshold, $M_{\rm D}=1.68 M_\odot$.
}
\label{fig6}      
\end{figure*}     

Figure \ref{fig6} 
is similar to Figs.\ \ref{fig2} 
and \ref{fig4} and shows cooling curves
calculated for the APR~III EOS and 
the same models for proton and 
neutron pairing (p and nt).
Neutron star masses  
range from 1.01 to $M_{\rm max}=1.93~M_\odot$. 
A sharp lower boundary of the densely covered domain
corresponds to the direct Urca threshold 
$M_{\rm D}=1.68 M_\odot$.  
In contrast to Fig.\ \ref{fig4}, we now obtain
a representative (densely populated) family
of middle-aged neutron stars with $M > M_{\rm D}$
and $\lg T_{\rm s}^\infty \approx 5.6-5.7$.
They are much colder than stars with $M<M_{\rm D}$
but much hotter than those with $M \approx M_{\rm max}$.  
They appear owing to the flattening of the dependence  
 $T_{\rm s}^\infty$ on $M$ by neutron superfluidity  
(Fig.\ \ref{fig5}). Were this cooling scenario realized
in the nature, the chances to detect such stars would be
higher than for the scenario in Fig.\ \ref{fig4}. 

The comparison of Figs.\ \ref{fig4} and \ref{fig6} 
shows a slight increase of 
the gap between the lower boundary 
$T_{\rm s}^\infty(t)$ of densely populated family
of neutron stars with $M<M_{\rm D}$
and the observed surface temperature  
of the Vela pulsar. 
This difference is also illustrated 
in Fig.\ \ref{fig5}; 
the curves $T_{\rm s}^\infty (M)$ 
and  $T_{\rm s}^\infty (\rho_{\rm c})$
merge in nearly the same slowly decreasing function
before the direct Urca thresholds
($M<M_{\rm D}$, $\rho_{\rm c}< \rho_{\rm D}$). 

\section{Discussion and Conclusions}

We have extended
the scenario of neutron star cooling proposed in Papers~I and II   
to the class of EOSs 
which allow the operation of the powerful
direct Urca process in the cores of massive stars. 
It widens possible interpretations of
observational data on thermal emission
from isolated middle-aged neutron stars 
using simplest models of neutron star cores 
composed of nucleons, electrons,
and muons. 

The main common feature of our present scenario and
the scenarios of Papers~I and II    
is that the enhanced neutrino emission,
required for the interpretation 
of observations of coldest neutron stars,
is produced by
Cooper pairing of neutrons. 
This puts stringent constraints (Paper~I) on the density
dependence of the critical temperature
$T_{\rm cnt}(\rho)$ for triplet-state neutron pairing.  
Tuning our phenomenological model
of $T_{\rm cnt}(\rho)$ 
we have obtained a noticeable dependence of
cooling on neutron star mass $M$. 

In the scenario of Paper~I we have also needed strong
proton pairing in the outer cores of neutron stars
to explain the observations of neutron stars hottest
for their age. 
In Paper~II we have extended that scenario
taking into account a possible presence of
accreted envelopes and 
singlet-state pairing of neutrons in the stellar crust. 
We have shown that the presence of accreted envelopes
allows us to take weaker proton pairing. 
Note, that models of weaker pairing are consistent
with recent  microscopic calculations    
of proton critical temperatures  
(\citealt{zuo04}, \citealt{tt04}), 
although some other calculations predict much
stronger proton pairing (e.g.,  \citealt{ls01};  
also see references in  \citealt{yls99}, 
and a recent paper by  \citealt{tmc04}).  
  
In this paper, for simplicity, we use the model of
strong proton pairing in the stellar core.
However, following Paper~II, 
we could assume  the presence 
of accreted envelopes, which 
would again allow us to weaken  
proton pairing.  
Adopting moderately strong  
proton pairing
(with $T_{\rm cp}^{\rm max} \ga 10^9$~K) 
we could interpret the 
observations of 
neutron stars warmest for their age.

The main feature of this paper is 
the consideration of EOSs  
which open 
the direct Urca process in central layers 
($\rho \ga 10^{15}$~g~cm$^{-3}$) 
of most massive  
stars. We have employed the parameterization
of the  EOS of \citet{apr98}
proposed by \citet{hhj99, hhj00}.  
In combination with properly chosen phenomenological models for
strong proton pairing 
($T_{\rm cp}^{\rm max} \ga 5 \times 10^9$~K
at $\rho \la  8 \times 10^{14}$~g~cm$^{-3}$) 
and moderate neutron pairing   
($T_{\rm cnt}^{\rm max} \sim 6 \times 10^8$~K
at $\rho \ga 10^{15}$~g~cm$^{-3}$)
it allows us to predict
new types of cooling neutron stars, in addition to
those given by the cooling scenario without
any direct Urca process (Papers~I and II, Sect.~3).
They are rapidly cooling neutron stars with 
masses higher than the direct Urca threshold mass 
$M_{\rm D}$. We have shown (Section 4) that the family
of these stars can be either rarely populated (Fig.\ \ref{fig4})
and thus difficult to discover, or rather
distinct and populated (Fig.\ \ref{fig6}), depending on the
choice of a model EOS of dense matter and a model for neutron
superfluidity. Neutron stars from the distinct class could be cold but
much warmer than massive stars with the fully open direct Urca
process. 

There is no clear evidence for the existence of
such cold stars at the moment but it may appear in the future.
For instance, low upper limits on thermal
emission from hypothetical neutron stars 
in a few supernova remnants 
\citep{kaplan04}  
can be treated as 
an indirect indication of
the existence of such cold stars. 
In this context    
the new scenario would allow one
to explain the existence of these cold objects
(if discovered) in the frame of standard and
simple physics of dense matter (without assuming
the presence of hyperons and/or exotic forms of matter,
such as pion/kaon condensates or quarks). 

As has already been emphasized, 
we need neutron pairing
to explain the current observations 
of stars coldest for their age.
However, according to
\citet{gkyg04b}, cooling curves 
are not too sensitive to 
exchanging
$T_{\rm cp}(\rho)  \rightleftharpoons T_{\rm cnt}(\rho)$
in neutron star cores. 
Therefore, we would also be able to
explain the data assuming
strong neutron pairing and 
moderate proton pairing
in stellar cores.  

Finally, let us note   
that the same physics of neutron star
interiors, which is tested by observations of isolated
(cooling) neutron stars, can also be tested   
by observations of accreting neutron stars in soft X-ray transients
(e.g., Yakovlev, Levenfish \& Haensel 2003) basing on the
hypothesis of deep crustal heating 
of such stars 
\citep{bbr98} by
pycnonuclear reactions in accreted matter \citep{hz90}. 
The observations of soft X-ray
transients in quiescent states indicate 
(Yakovlev, Levenfish \& Gnedin 2005) the existence of 
rather cold neutron stars (first of all, SAX J1808.4--3658)
inconsistent with the minimal model 
of cooling neutron stars.  
Although these observations are currently 
inconclusive (e.g., \citealt{ylg05}),
if confirmed in future observations, they could give
an evidence in favor 
of the cooling scenario
proposed in this paper.


\noindent
\textit{Acknowledgments}
We are grateful to A.Y.\ Potekhin for providing us with
fit parameters for the APR EOS, as well as to A.M.\ Krassilchtchikov
and K.P.\ Levenfish for assistance in drawing figures.
We are also grateful to anonymous referee for critical remarks.
This work has been supported partly by
the RFBR (grants 05-02-16245 and 03-07-90200),
the program Russian Leading Scientific School (grant 1115.2003.2),
the Russian Science Support Foundation,
and by the INTAS (grant YSF 03-55-2397).  \\


\end{document}